\documentclass{llncs}

\usepackage{caption}
\usepackage{ifthen}
\usepackage{listings}
\usepackage{amsmath}
\usepackage{color}
\usepackage{graphicx} 
\usepackage{colortbl}
\usepackage{float}
\usepackage{xspace}
\usepackage{multirow}

\let\othelstnumber=\thelstnumber
\def\createlinenumber#1#2{
    \edef\thelstnumber{%
        \unexpanded{%
            \ifnum#1=\value{lstnumber}\relax
              #2%
            \else}%
        \expandafter\unexpanded\expandafter{\thelstnumber\othelstnumber\fi}%
    }
    \ifx\othelstnumber=\relax\else
      \let\othelstnumber\relax
    \fi
}

\newcommand{\figpath}{./}

\newcommand{\level}{level\xspace}

\newcommand{\levels}{levels\xspace}
\newcommand{\hcir}{HC IR\xspace}
\newcommand{\llvmir}{LLVM IR\xspace}
\newcommand{\llvm}{LLVM\xspace}
\newcommand{\isa}{ISA\xspace}
\newcommand{\ciao}{Ciao\xspace}
\newcommand{\ciaopp}{CiaoPP\xspace}
\newtheorem{auxexample}{Example}[section]

\definecolor{lightgrey}{rgb}{0.95,0.95,0.95}

\newcommand{\tabend}{\vspace*{-5mm}}
\newcommand{\figend}{\vspace*{-5mm}}

\def\imp{\hbox{${\tt \ :\!-\ }$}}

\lstdefinelanguage{xc}{
  keywords={int,if,return, struct, void},
  keywordstyle=\bf,
  basicstyle=\scriptsize
}[keywords]

\lstdefinelanguage{ciao}{
  keywords={regtype, trust, num, list, rsize, var, resource},
  keywordstyle=\bf,
  basicstyle=\scriptsize,
}[keywords]

\lstdefinelanguage{llvm}{
  keywords={getelementptr, label, br, i1, i32, ret, phi, ne, icmp, sub},
  basicstyle=\scriptsize,	
  keywordstyle=\bf,
  xleftmargin=.02\textwidth,
  frame=l,
  numbers=left,
  numberstyle=\scriptsize
}[keywords]

\lstdefinelanguage{hcir}{
  keywords={trust},
  basicstyle=\scriptsize,
  keywordstyle=\bf,
  frame=l,
  xleftmargin=.02\textwidth, %xrightmargin=.05\textwidth,
  numbers=left,
  numberstyle=\scriptsize
}[keywords]

\definecolor{darkgreen}{rgb}{0,0.75,0}
\definecolor{darkblue}{rgb}{0,0,0.75}
\definecolor{lightgreen}{rgb}{0.85,1,0.85}
\definecolor{lightblue}{rgb}{0.85,0.85,1}
\definecolor{grey85}{rgb}{0.45,0.45,0.45}
\definecolor{lightblue2}{rgb}{0.75,0.75,1}

\definecolor{lightblue3}{rgb}{0.70,0.70,0.92}
\definecolor{magenta}{rgb}{1,0,1}
\definecolor{sienna}{rgb}{1,0,1}
\definecolor{maroon4}{rgb}{0.55,0.11,0.38}
\definecolor{chocolate}{rgb}{0.82,0.41,0.12}
\definecolor{red4}{rgb}{0.55,0,0}
\definecolor{darkred}{rgb}{0.75,0,0}

\title{Inferring Parametric Energy Consumption Functions at Different Software Levels: \\ISA vs.\ \llvmir} 

\author{U.~Liqat\inst{1} \and K.~Georgiou\inst{2} \and 
   S.~Kerrison\inst{2} \and P.~Lopez-Garcia\inst{1,3} \and
   John P. Gallagher\inst{5} \and M.V.~Hermenegildo\inst{1,4}  \and K.~Eder\inst{2} }

\institute{
IMDEA Software Institute, Madrid, Spain \\
\email{\{umer.liqat,pedro.lopez,manuel.hermenegildo\}@imdea.org}
\and 
University of Bristol, Bristol, UK \\
\email{\{kyriakos.georgiou,steve.kerrison,kerstin.eder\}@bristol.ac.uk}
\and 
Spanish Council for Scientific Research (CSIC), Madrid, Spain \and 
Universidad Polit\'{e}cnica de Madrid (UPM), Madrid, Spain \and
Roskilde University, Roskilde, Denmark\\
\email{jpg@ruc.dk}
}

\begin{document}

\maketitle

\begin{abstract}

The static estimation of the energy consumed by program executions is
an important challenge, which has applications in program optimization
and verification, and is instrumental in energy-aware software
development.  Our objective is to estimate such energy consumption in
the form of \emph{functions on the input data sizes of programs}.
We have developed a tool for experimentation with static analysis
which infers such energy functions at two levels, the instruction set
architecture (ISA) and the intermediate code (\llvmir) levels, and
reflects it upwards to the higher source code level. This required the
development of a translation from \llvmir to an intermediate
representation and its integration with existing components, a
translation from ISA to the same representation, a resource analyzer,
an ISA-level energy model, and a mapping from this model to \llvmir.
The approach has been applied to programs written in the XC language
running on XCore architectures, but is general enough to be applied to
other languages.
Experimental results show that our \llvmir \level analysis is
reasonably accurate 
(less than $6.4\%$ average error vs.\ hardware measurements) and more
powerful than analysis at the ISA \level. This paper provides 
insights into the trade-off 
of precision versus analyzability at these levels.


\end{abstract}

\section{Introduction}
\label{introduction}

Energy consumption and the environmental impact of computing
technologies have become a major worldwide concern.  It is an important
issue in high-performance computing, distributed applications, and
data centers.  There is also increased demand for complex computing
systems which have to operate on batteries, such as
implantable/portable medical devices or mobile phones.
Despite advances in power-efficient hardware, more energy savings can
be achieved by improving the way current software technologies make
use of such hardware.

The process of developing energy-efficient software can benefit
greatly from static analyses that estimate the energy consumed by
program executions without actually running them.  Such estimations
can be used for different software-development tasks, such as
performing automatic optimizations, verifying energy-related
specifications, and helping system developers to better understand the
impact of their designs on energy consumption. These tasks often
relate to the source code \level.
% Extension
For example, source-to-source transformations to produce optimized
programs are quite common. Specifications included in the source code
can be proved or disproved by comparing them with safe information
inferred by analysis. Such information, when referred to the
procedures in the source code can be useful for example to detect
which are the most energy-consuming ones and replace them by more
energy-efficient implementations.
On the other hand, energy consumption analysis must typically be
performed at lower levels in order to take into account the effect of
compiler optimizations and to link to an energy model. Thus, the inference of energy consumption
information for lower \levels such as the Instruction Set Architecture
(ISA) or intermediate compiler representations (such as
\llvmir~\cite{LattnerLLVM2004}) is fundamental for two reasons:
1) It is an intermediate step that allows propagation of energy
consumption information from such lower \levels up to the source code
\level;
and 2) it enables optimizations or other applications at the ISA and
\llvmir \levels.  

In this paper (an improved version
of~\cite{entra-d3.2.4-isa-vs-llvm-short}) we propose a static analysis
approach
that infers energy consumption information at the ISA and \llvmir
\levels, and reflects it up to the source code \level.  Such information
is provided in the form of \emph{functions on input data sizes}, and
is expressed by means of \emph{assertions} that are inserted in the
program representation at each of these \levels.
The user (i.e., the ``energy-efficient software developer'') can
customize the system by selecting the \level at which the analysis
will be performed (ISA or \llvmir) and the \level at which energy
information will be output (ISA, \llvmir or source code).
As we will show later, the selection of analysis level has an impact
on the analysis accuracy and on the class of programs that can be
analyzed.

The main goal of this paper is to study the feasibility and
practicability of the proposed analysis approach and perform an
initial experimental assessment to shed light on the trade-offs implied
by performing the analysis at the ISA or \llvm \levels.
In our experiments
we focus on the energy analysis of programs written in
XC~\cite{Watt2009}
running on the XMOS XS1-L architecture.  However, the concepts
presented here are neither language nor architecture dependent and
thus can be applied to the analysis of other programming languages
(and associated lower \level program representations) and
architectures as well.
XC is a high-level C-based programming language that includes
extensions for concurrency, communication, input/output operations,
and real-time behavior.   
In order to potentially support different programming languages and
different program representations at different \levels of compilation
(e.g., \llvmir and ISA)
in the same analysis framework 
we differentiate between the \emph{input language} (which can be XC
source, \llvmir, or ISA) and the \emph{intermediate semantic program
  representation} that the resource analysis operates on. The latter
is a series of connected code blocks, represented by Horn Clauses,
that we will refer to as ``\hcir'' from now on.
We then propose a transformation from each \emph{input language} into
the \hcir and passing it to a resource analyzer. 
The \hcir representation as well as a transformation from \llvmir into \hcir will be 
explained in Section~\ref{sec:llvm-ciao-translation}.
In our implementation we use an extension of the
\ciaopp~\cite{ciaopp-sas03-journal-scp} resource analyzer. This
analyzer always deals with the \hcir  in the same way, independent of
its origin,
inferring energy consumption functions for all procedures in the \hcir
program.
The main reason for choosing Horn Clauses as the intermediate
representation is that it offers a good number of features that make
it very convenient for the
analysis~\cite{decomp-oo-prolog-lopstr07}. For instance, it supports
naturally Static Single Assignment (SSA) and recursive forms, 
as will be explained later.  
In fact, there is a current trend favoring the use of Horn Clause
programs as intermediate representations in analysis and verification
tools~\cite{DBLP:conf/tacas/GrebenshchikovGLPR12,DBLP:conf/fm/HojjatKGIKR12-short,z3,hcvs14}.

Although our experiments are based on single-threaded XC programs
(which do not use pointers, since XC does not support them), our claim
about the generality and feasibility of our proposed approach for
static resource analysis is supported by existing tools based on the
Horn Clause representation that can successfully deal with C source
programs that exhibit interesting features such as the use of
pointers, arrays, shared-memory, or concurrency in order to analyze
and verify a wide range of
properties~\cite{DBLP:conf/tacas/GrebenshchikovGLPR12,DBLP:conf/fm/HojjatKGIKR12-short,DBLP:conf/tacas/GurfinkelKN15-short}. For
example~\cite{DBLP:conf/tacas/GurfinkelKN15-short} is a tool for the
verification of safety properties of C programs which can reason about
scalars and pointer addresses, as well as memory contents. It
represents the bytecode corresponding to a C program by using
(constraint) Horn clauses.

Both static analysis and energy models can potentially relate to any language \level (such as 
XC source, \llvmir, or ISA).
Performing the analysis 
at a
given \level means that 
the representation of the program at
that \level 
is transformed into the \hcir, and 
the analyzer
``mimics'' the semantics of instructions at that level.
The energy
model at a given \level provides basic information on the energy cost of instructions
at that level.
The analysis results at a given \level can be mapped upwards to
a higher level, e.g. from ISA or \llvmir to XC.
Furthermore, it is possible to perform analysis at a given level
with an energy model for a lower level. In this case the energy model
must 
be reflected up to the analysis level.

Our hypothesis is that 
the choice of \level will affect the accuracy of the energy models and
the precision of the analysis in opposite ways: energy models at lower
\levels (e.g. at the ISA \level) will be more precise than at higher
\levels (e.g. XC source code), since the closer to the hardware, the
easier it is to determine the effect of the execution on the hardware.
However, at lower \levels more program structure and data type/shape
information is lost due to lower-level representations, and we expect
a corresponding loss of analysis accuracy. 
% (without using complex techniques for recovering type
% information and abstracting memory operations).
% Extension
We could devise mechanisms to represent such higher-level information
and pass it down to the lower-level \isa, or to recover it by
analysing the \isa. However, our goal is to compare the analysis at
the \llvmir and \isa \levels without introducing such mechanisms,
which might be complex or not effective in some cases (e.g., in
abstracting memory operations or recovering type information).
% End

\begin{figure}[ht]
\centering
\centerline{\includegraphics[scale=0.45]{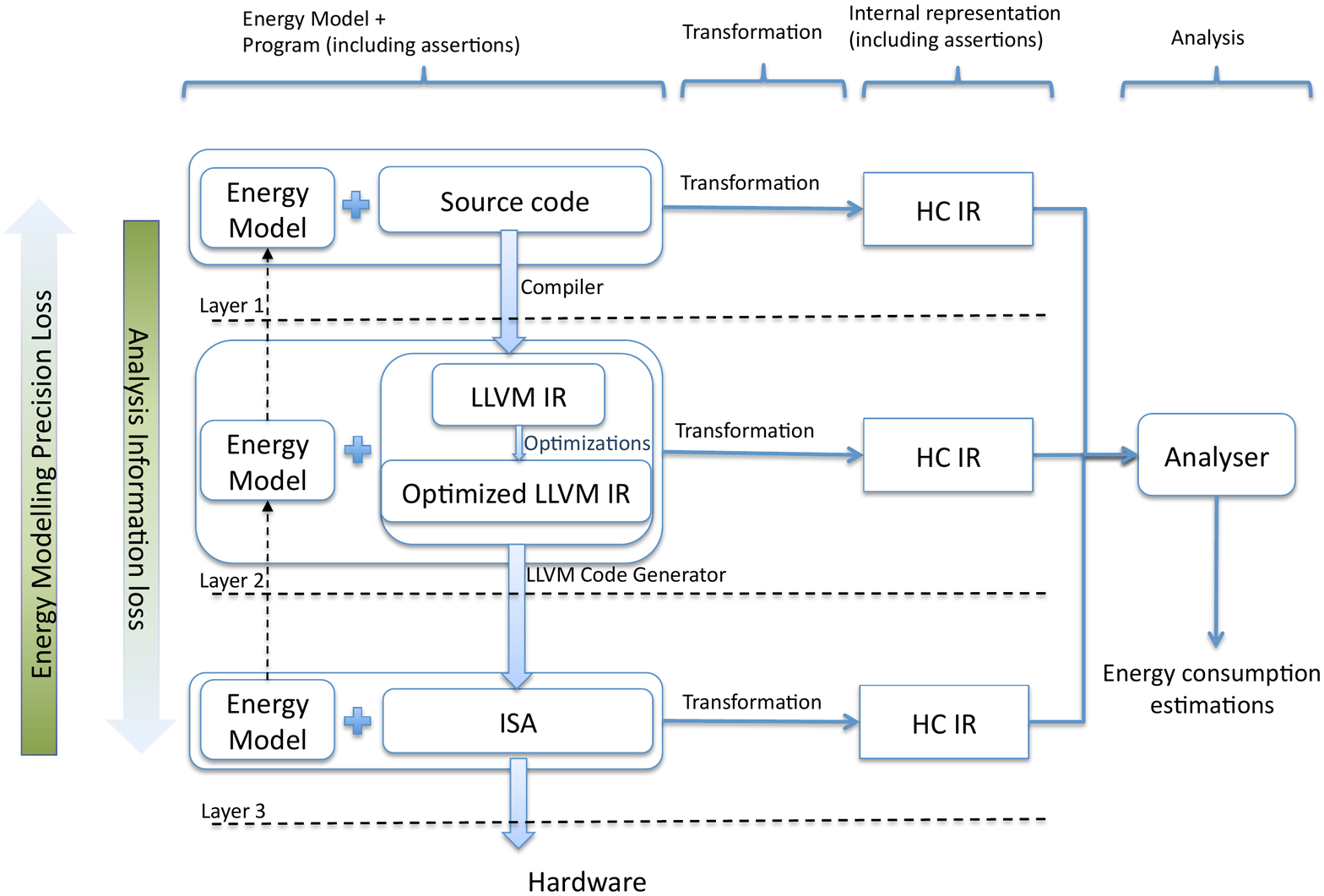}}
\vspace{-0.7cm}
\caption{Analysis/modelling \level trade-off and potential choices.}
\label{analysis-model-choices-overview}
\end{figure}

This hypothesis about the analysis/modelling \level trade-off (and
potential choices) is illustrated in
Figure~\ref{analysis-model-choices-overview}.
The possible choices are classified into two groups: those that
analyze and model at the same \level, and those that operate at
different \levels. For the latter, the problem is finding good
mappings between software segments from the \level at which the model
is defined up to the \level at which the analysis is performed, in a
way that does not lose accuracy in the energy information.

In this paper
we concentrate on two of these choices and their comparison, to see if
our hypothesis holds.
In particular, the first approach (choice~1) is represented by
analysing the generated ISA-\level code using 
models defined at the ISA \level that express the energy consumed by
the execution of individual ISA instructions.
This approach was explored
in~\cite{isa-energy-lopstr13-final}.  It used the precise ISA-\level
energy models presented in~\cite{Kerrison13}, which when used in the
static analysis of~\cite{isa-energy-lopstr13-final} for a number of
small numerical programs resulted in the inference of functions that provide
reasonably accurate energy consumption estimations for any input data
size (3.9\% average error vs.\ hardware measurements).
However, when dealing with programs involving structured types such as arrays, it
also pointed out
that, due to the loss of information related to program structure and
types of arguments at the ISA \level (since it is compiled away and no
longer relates cleanly to source code), the power of the analysis was
limited.
In this paper we start by exploring an alternative approach: the
analysis of the generated \llvmir
(which retains much more of such information, enabling more direct
analysis as well as mapping of the analysis information back to source
\level)
together with techniques that map segments of ISA instructions to \llvmir
blocks~\cite{Georgiou2015arXiv} (choice~2). This mapping is used to propagate
the energy model information defined at the ISA \level up to the
\level at which the analysis is performed, the \llvmir \level.  In
order to complete the \llvmir-\level analysis, we have also developed
and implemented a transformation from \llvmir into \hcir 
and used the \ciaopp resource analyzer. This results in a parametric
analysis that similarly to~\cite{isa-energy-lopstr13-final} infers
energy consumption functions, but operating on the \llvmir \level
rather than the ISA \level.  

We have performed an experimental comparison of the two choices
for generating energy consumption functions. Our results support
our intuitions about the trade-offs involved. They also provide
evidence that the \llvmir-\level analysis (choice~2) offers a good
compromise within the \level hierarchy, since it broadens the class of
programs that can be analyzed 
% Extension
without the need for developing complex techniques for recovering type
information and abstracting memory operations, and
% End
without significant loss of accuracy.

In summary, the original contributions of this paper are:
\begin{enumerate}

\item A translation from \llvmir to \hcir
  (Section~\ref{sec:llvm-ciao-translation}).

\item The integration of all components into an experimental tool
  architecture, enabling the static inference of energy consumption
  information in the form of \emph{functions on input data sizes} and
  the experimentation with the trade-offs described above
  (Section~\ref{sec:overview}). The components are: \llvmir and \isa
  translations, \isa-\level energy model and
  mapping technique
  (Section~\ref{sec:energy-model} and
  ~\cite{Kerrison13,Georgiou2015arXiv}),
  and analysis tools (Section~\ref{sec:ciaopp-analysis}
  and~\cite{resource-iclp07,plai-resources-iclp14}).

\item The experimental results and evidence of trade-off of precision
  versus analyzability (Section~\ref{sec:experiments}).

\item A sketch of how the static analysis system can be integrated in
  a source-level Integrated Development Environment (IDE)
  (Section~\ref{sec:overview}).

\end{enumerate}

Finally, some related work is discussed in
Section~\ref{sec:related-work}, and Section~\ref{sec:conclusions}
summarises our conclusions and comments on ongoing and future work.

\section{Overview of the Analysis at the \llvmir Level} % Here macro does not work (capitals!)
\label{sec:overview}

\begin{figure}[ht]
\centering
\includegraphics[scale=0.45]{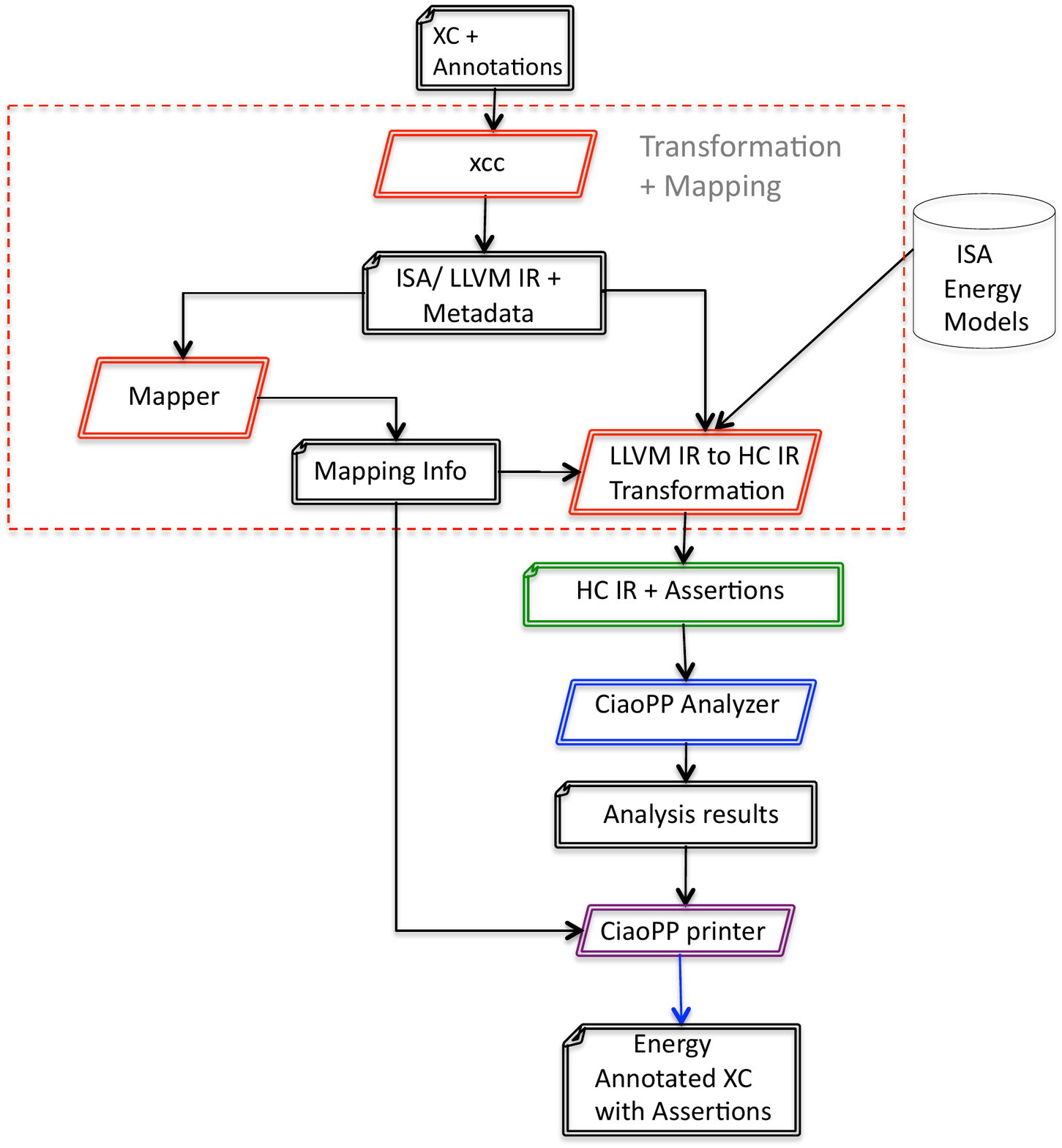}
 \vspace{-1cm}
\caption{An overview of the analysis at the \llvmir \level using ISA models.}
\figend
\label{fig:ciaopp-framework-llvm}
\end{figure}

An overview of the proposed analysis system at the \llvmir \level
using models at the ISA \level is depicted in
Figure~\ref{fig:ciaopp-framework-llvm}.
The system takes as input an XC source program that can (optionally)
contain assertions (used to provide useful hints and information to
the analyzer), from which a \emph{Transformation and Mapping} process
(dotted red box) generates first its associated \llvmir using the xcc
compiler. Then, a transformation from \llvmir into \hcir is performed
(explained in Section~\ref{sec:llvm-ciao-translation}) obtaining the
intermediate representation (green box) that is supplied to the
\ciaopp analyzer. This representation includes assertions that express
the energy consumed by the \llvmir blocks, generated from the
information produced by the mapper tool (as explained in
Section~\ref{sec:energy-model}). The \emph{\ciaopp analyzer} (blue
box, described in Section~\ref{sec:ciaopp-analysis}) takes the \hcir,
together with the assertions which express the energy consumed by
\llvmir blocks, and possibly some additional (trusted) information,
and processes them, producing the analysis results, which are
expressed also using assertions.
Based on the procedural interpretation of these \hcir programs and 
the resource-related information contained in the assertions, the resource
analysis can infer static bounds on the energy
consumption of the \hcir programs that are applicable to the original
\llvmir and, hence, to their corresponding XC programs.
The analysis results include energy consumption information expressed
as functions on data sizes for the whole program and for all the
procedures and functions in it. Such results are then processed by the
\emph{\ciaopp printer} (purple box) which presents the
information to the program developer in a user-friendly format.

\section{\llvmir to \hcir Transformation}
\label{sec:llvm-ciao-translation}
In this section we describe the \llvmir to \hcir transformation that
we have developed in order to achieve the complete analysis system
at the \llvmir \level proposed in the paper (as already mentioned in the
overview given in Section~\ref{sec:overview} and depicted in
Figure~\ref{fig:ciaopp-framework-llvm}).

A Horn clause (HC) is a first-order predicate logic formula of the form $\forall (S_1 \wedge \ldots \wedge S_n \rightarrow S_0)$ where all variables in the clause are universally quantified over the whole formula, and $S_0,S_1,\ldots,S_n$ are atomic formulas, also called literals.  It is usually written $S_0 \imp S_1,\ldots,S_n$.

The \hcir representation consists of a sequence of \emph{blocks} where
each block is represented as a \emph{Horn clause}:

\centerline{\mbox{$<block\_id>( <params> ) \imp \ S_1, \ \ldots \ ,S_n.$}}

\noindent
Each block has an entry point, that we call the \emph{head} of the
block (to the left of the $\imp$ symbol),
with a number of parameters $<params>$, and a sequence of steps (the
\emph{body}, to the right of the $\imp$ symbol). Each of these $S_i$
steps (or \emph{literals}) is either (the representation of) an
\llvmir \emph{instruction}, or a \emph{call} to another (or the same)
block.
% Extension
The analyzer deals with the \hcir always in the same way, independent
of its origin.
% End
The transformation ensures that the program information relevant to
resource usage is preserved, so that the energy consumption functions
of the \hcir programs inferred by the resource analysis are applicable
to the original \llvmir programs.

The transformation also 
passes energy values for the
\llvmir \level for different programs based on the ISA/\llvmir mapping
information that express the energy consumed by the \llvmir blocks, as
explained in Section~\ref{sec:energy-model}. Such information is
represented by means of \emph{trust} assertions (in the \ciao
assertion language~\cite{hermenegildo11:ciao-design-tplp-short}) that
are included in the \hcir.
In general, \emph{trust} assertions can be used to provide information
about the program and its constituent parts (e.g., individual
instructions or whole procedures or functions) to be trusted by the
analysis system, i.e., they provide base information assumed to be
true by the inference mechanism of the analysis in order to propagate
it throughout the program and obtain information for the rest of its
constituent parts.

\llvmir programs are expressed using typed assembly-like
instructions. Each function is in SSA form, represented as a sequence
of basic blocks. Each basic block is a sequence of \llvmir
instructions that are guaranteed to be executed in the same order.
Each block ends in either a branching or a 
return instruction. 
In order to transform an \llvmir program into the \hcir, 
we follow a similar approach as in a 
previous
ISA-\level transformation~\cite{isa-energy-lopstr13-final}. 
However, the \llvmir includes an additional type transformation as
well as better memory modelling. 

The following subsections describe the main aspects of the transformation. 
\subsection{Inferring Block Arguments} 
As described before, a \textit{block} in the \hcir has an entry point
(head) with input/output parameters, and a body containing a sequence
of steps
(here, representations of \llvmir instructions). 
Since the scope of the variables in \llvmir blocks is at the function
level, the blocks are not required to pass parameters while making
jumps to other blocks. Thus, in order to represent \llvmir blocks as
\hcir blocks, we need to infer input/output parameters for each block.

For entry blocks, the input and output arguments are the same as the ones to the function.
We define the functions
$param_{in}$ and $param_{out}$ which infer input and output parameters
to a block respectively. 
These are recomputed according to the following definitions until a
fixpoint is reached:
\begin{align*}
\mathit{params}_{out}(b) & = \textstyle (kill(b) \ \cup\ \mathit{params}_{in}(b))
      \ \cap\ \bigcup_{b'\in \mathit{next}(b)}\mathit{params}_{out}(b') \\
\mathit{params}_{in}(b) & = \textstyle gen(b)\ \cup\ \bigcup_{b'\in \mathit{next}(b)} \mathit{params}_{in}(b')
\end{align*}
where $\mathit{next}(b)$ denotes the set of immediate target blocks
that can be reached from block $b$ with a jump instruction, while $gen(b)$
and $kill(b)$ are the read and written variables in block $b$
respectively, which are defined as:
\begin{align*}
kill(b)&=\textstyle\bigcup\limits_{k=1}^{n}\mathit{def}(k) \\
gen(b)&=\textstyle\bigcup\limits_{k=1}^{n}\{v\mid v\in \mathit{ref}(k) \wedge \forall (j<k). v\notin \mathit{def}(j)\}
\end{align*}
where $\mathit{def}(k)$ and $\mathit{ref}(k)$ denote the variables
written or referred to at a node (instruction) $k$ in the block,
respectively, and $n$ is the number of nodes in the block.

Note that the \llvmir is in SSA form at the function level, which
means that blocks may have $\phi$ nodes which are created while
transforming the program into SSA form. A $\phi$ node is essentially a
function defining a new variable by selecting one of the multiple
instances of the same variable coming from multiple predecessor
blocks:
\begin{center}
$x=\phi(x_1, x_2, ..., x_n)$
\end{center}
$\mathit{def}$ and $\mathit{ref}$ for this instruction are $\{x\}$ and $\{x_1, x_2, ... ,
x_n\}$ respectively. 
An interesting feature of our approach is that $\phi$ nodes are not
needed. Once the input/output parameters are inferred for each block
as explained above, a post-process gets rid of all $\phi$ nodes by
modifying block input arguments in such a way that blocks receive $x$
directly as an input and an appropriate $x_i$ is passed by the call
site. This will be illustrated later in
Section~\ref{sec:llvm2isa-block-trans}.

Consider the example in Figure~\ref{llvm2hcirblock} (left), where the \llvmir
block \textit{looptest} is defined. The body of the block reads from 2
variables without previously defining them in the same block. The
fixpoint analysis would yield:
\begin{center}
$params_{in}(looptest) = \{Arr, I\}$
\end{center}
which is used to construct the \hcir representation of the \textit{looptest}
block shown in Figure~\ref{llvm2hcirblock} (right), line 3.

\subsection{Translating \llvmir Types into \hcir Types} 

\llvmir is a typed representation which allows retaining much more of
the (source) program information than the ISA representation (e.g.,
types defining compound data structures).
% Extended
As already mentioned, this enables a more direct analysis as well as
mapping of the analysis information back to source level.
% End
Thus, we define a mechanism to translate \llvmir types into their
counterparts in \hcir.

The LLVM type system defines primitive and derived types. The
primitive types are the fundamental building blocks of the type
system. Primitive types include \textit{label, void, integer,
  character, floating point, x86mmx,} and \textit{metadata}. The
\textit{x86mmx} type represents a value held in an MMX register on an
x86 machine and the \textit{metadata} type represents embedded
metadata. The derived types are created from primitive types or other
derived types. They include \textit{array, function, pointer,
  structure, vector, opaque}. Since the XCore platform 
supports neither pointers nor floating point data types, the \llvmir code
generated from XC programs uses only a subset of the LLVM types.

At the \hcir level we use \emph{regular types}, one of the type
systems supported by \ciaopp~\cite{ciaopp-sas03-journal-scp}.  Translating \llvmir primitive types
into regular types is straightforward. The \textit{integer} and
\textit{character} types are 
abstracted as \textit{num} regular type, whereas the
\textit{label}, \textit{void}, and \textit{metadata} types are
represented as \textit{atm} (atoms). 

For derived types, corresponding non-primitive regular types are
constructed during the transformation phase. Supporting non-primitive
types is important because it enables the analysis to infer energy
consumption functions that depend on the sizes of internal parts of
complex data structures. 
The array, vector, and structure types are represented as
follows:
\begin{center}
$array\_type\rightarrow (nested) list$\\
$vector\_type\rightarrow (nested) list$\\
$structure\_type\rightarrow functor\_term$ 
\end{center}
Both the \textit{array} and \textit{vector} types are represented by
the \textit{list} type in \ciaopp which is a special case of compound
term. The type of the elements of such lists can be again a primitive
or a derived type. The \textit{structure} type is represented by a
compound term which is composed of an atom (called the
\textit{functor}, which gives a name to the structure) and a number of
\textit{arguments}, which are again either primitive or derived
types. LLVM also introduces pointer types in the intermediate
representation, even if the front-end language does not support them
(as in the case of XC, as mentioned before). Pointers are used in the
pass-by-reference mechanism for arguments, in memory allocations in
\textit{alloca} blocks, and in memory load and store operations. The
types of these pointer variables in the \hcir are the same as the
types of the data these pointers point to.

\begin{figure}[ht]

\begin{tabular}{  c | c  }
\begin{lstlisting}[language=xc]
struct mystruct{
  int x;
  int arr[5];
};

void print(struct mystruct [] Arg, int N)
{ 
  ...
}
\end{lstlisting}

&

\begin{lstlisting}[language=ciao]
:- regtype array1/1. 
array1:=[] | [~struct|array1].

:- regtype struct/1.

struct:=mystruct(~num,~array2).

:- regtype array2/1.
array2:=[] | [~num|array2].
\end{lstlisting}
\end{tabular}

\caption{An XC program and its type transformation into \hcir.}
\label{xc2regtypes}
\figend
\end{figure}

Consider for example the types in the XC program shown in
Figure~\ref{xc2regtypes}.  The type of argument $Arg$ of the $print$
function is an array of $mystruct$ elements. $mystruct$ is further
composed of an integer and an array of integers.  The \llvmir
code generated by xcc for the function signature $print$ in Figure~\ref{xc2regtypes} (left) is:
\vspace{-1mm}
\begin{center}
define void @print( $[0 \times \{ i32, [5 \times i32] \}]$* noalias nocapture)
\end{center}
\vspace{-1mm}
The function argument type in the \llvmir ($[0 \times \{ i32, [5
\times i32] \}]$) is the typed representation of the argument $Arg$ to
the function in the XC program. It represents an array of arbitrary length with 
 elements of $\{ i32, [5 \times i32] \}$ structure type which is further
composed of an $i32$ integer type and a $[5 \times i32]$ array type,
i.e., an array of 5 elements of $i32$ integer type.\footnote{$[0 \times i32]$ specifies an arbitrary length array of $i32$ integer type elements. }

This type is represented in the \hcir using the set of regular types
illustrated in Figure~\ref{xc2regtypes} (right). The regular type $array1$,
is a list of $struct$ elements (which can also be simply written as
\texttt{array1 := list(struct)}). Each $struct$ type element is
represented as a functor $mystruct/2$ where the first argument is a
$num$ and the second is another list type $array2$.  The type $array2$
is defined to be a list of $num$ (which, again, can also be simply
written as \texttt{array2 := list(num)}).

\subsection{Transforming \llvmir Blocks/Instructions into \hcir}
\label{sec:llvm2isa-block-trans}

In order to represent an \llvmir function by an \hcir
function (i.e., a predicate), we need to represent each \llvmir block by an \hcir block
(i.e., a Horn clause) and hence each \llvmir instruction by an \hcir
literal.

\begin{figure}[ht]
\begin{minipage}[t]{0.04\textwidth}
\ 
\end{minipage}
\begin{minipage}[t]{0.51\textwidth}
\begin{lstlisting}[language=llvm]
alloca:
 br label looptest
looptest:
 %I=phi i32[%N,%alloca],             [%I1,%loopbody]
 %Zcmp=icmp ne i32 %I, 0
 br i1 %Zcmp, label %loopbody,          label %loopend
loopbody:
 %Elm=getelementptr [0xi32]*%Arr,               i32 0,i32 %I
 //process array element `Elm'
 %I1=sub i32 %I, 1
 br label %looptest
loopend:
 ret void
\end{lstlisting}
\end{minipage}
\begin{minipage}[t]{0.45\textwidth}
\begin{lstlisting}[language=hcir]
alloca(N, Arr):-
 looptest(N, Arr).
looptest(I, Arr):-
 icmp_ne(I, 0, Zcmp),
 loopbody_loopend(Zcmp,I,Arr).
icmp_ne(X, Y, 1):- X \= Y.
icmp_ne(X, Y, 0):- X = Y.
loopbody_loopend(Zcmp,I,Arr):-
 Zcmp=1,
 nth(I, Arr, Elm),
 //process list element `Elm'
 I1 is I - 1, sub(I,1,I1),
 looptest(I1, Arr).
loopbody_loopend(Zcmp,I,Arr):-
 Zcmp=0. 
\end{lstlisting}
\end{minipage}
\caption{\llvmir Array traversal example (left) and its \hcir representation (right)}
\label{llvm2hcirblock}
\figend
\end{figure}

The \llvmir{} instructions are transformed into equivalent \hcir{} literals where the 
semantics of the execution of the \llvmir{} instructions are either described using trust assertions or by giving definition to \hcir{} literals. The \textit{phi assignment} instructions are removed and the semantics of the \textit{phi assignment} are preserved on the call sites. For example, the \emph{phi assignment}
is removed from the \hcir block in Figure~\ref{llvm2hcirblock} (right) and the semantics of the \emph{phi assignment}
is preserved on the call sites of the \textit{looptest}
(lines 2 and 14).  The call sites $alloca$ (line 2) and $loopbody$ (line 13) pass the
corresponding value as an argument to $looptest$, which is received by $looptest$ in
its first argument $I$.  

Consider the instruction \textit{getelementptr} at line 8
in Figure~\ref{llvm2hcirblock} (left), which computes 
the address of an element of an array
\textit{\%Arr} indexed by \textit{\%I} and assigns it to a variable
\textit{\%Elm}. 
Such an instruction is represented by a call to an
abstract predicate \textit{nth/3}, which extracts a reference to an
element from a list, and whose effect of execution on energy 
consumption as well as the relationship between the sizes of input and output arguments is described using trust assertions. For example, the assertion:

\begin{lstlisting}[language=ciao, basicstyle=\scriptsize\ttfamily,]
:- trust pred nth(I, L, Elem)
  :(num(I), list(L, num), var(Elem)) 
  => ( num(I), list(L, num), num(Elem), 
       rsize(I, num(IL, IU)), 
       rsize(L, list(LL, LU, num(EL, EU))), 
       rsize(Elem, num(EL, EU)) )
       + (resource(avg, energy, 1215439) ).
\end{lstlisting}

\noindent 
\noindent
indicates that if the \texttt{nth(I, L, Elem)} predicate (representing
the \textit{getelementptr} \llvmir instruction) is called with
\textit{I} and \textit{L} bound to an integer and a list of numbers respectively,
and \textit{Elem} an unbound variable (precondition field
``\texttt{:}''), then, after the successful completion of the call
(postcondition field ``\texttt{=>}''), \textit{Elem} is an integer
number and the lower and upper bounds on its size are equal to the
lower and upper bounds on the sizes of the elements of the list
\textit{L}. The sizes of the arguments to \textit{nth/3} are expressed
using the property \textit{rsize} in the assertion language. The lower
and upper bounds on the length of the list \textit{L} are \textit{LL}
and \textit{LU} respectively. Similarly, the lower and upper bounds on
the elements of the list are \textit{EL} and \textit{EU} respectively,
which are also the bounds for \textit{Elem}. The
\textit{resource} property (global computational properties field “+”)
expresses that the energy consumption for the instruction
is an average value (1215439 nano-joules\footnote{nJ, $10^{-9}$~joules}).

The branching instructions in \llvmir{} are transformed into calls to target blocks in \hcir{}. For example, the branching instruction at line 6 in Figure~\ref{llvm2hcirblock} (left),
which jumps to one of the two blocks \textit{loopbody} or
\textit{loopend} based on the Boolean variable \textit{Zcmp}, is
transformed into a call to a predicate with two clauses (line 5 in Figure~\ref{llvm2hcirblock} (right)). The name of
the predicate is the concatenation of the names of the two \llvmir
blocks mentioned above. The two clauses of the predicate defined at lines 8-13 and 14-15 in
Figure~\ref{llvm2hcirblock} (right) represent the \llvmir blocks
\textit{loopbody} and \textit{loopend} respectively. The test on the
conditional variable is placed in both clauses to preserve the 
semantics of the conditional branch.

\section{Obtaining the Energy Consumption of \llvmir Blocks}
\label{sec:energy-model}

Our approach requires producing assertions that express the energy
consumed by each call to an \llvmir block (or parts of it) when it is
executed. To achieve this we take as starting point the energy
consumption information available from an existing XS1-L ISA Energy
Model produced in our previous work of ISA level analysis~\cite{isa-energy-lopstr13-final}
using the techniques described in~\cite{Kerrison13}. 
We refer the reader to~\cite{Kerrison13} for a detailed study
of the energy consumption behaviour of the XS1-L architecture,
containing a description of the test and measurement process along
with the construction and full evaluation of such model. In the
experiments performed in this paper a single, constant energy value is
assigned to each instruction in the ISA based on this model.

A mechanism is then needed to propagate such ISA-\level energy
information up to the \llvmir \level and obtain energy values for
\llvmir blocks. A set of mapping techniques serve this purpose by
creating a fine-grained mapping between segments of ISA instructions
and \llvmir code segments, in order to enable the energy
characterization of each \llvmir instruction in a program, by
aggregating the energy consumption of the ISA instructions mapped to
it. Then, the energy value assigned to each \llvmir block is obtained
by aggregating the energy consumption of all its \llvmir
instructions. The mapping is done by using the debug mechanism 
where the debug information, preserved during the lowering phase 
of the compilation from \llvmir to ISA, is used to track ISA instructions 
against \llvmir instructions. A full description and formalization of the mapping techniques 
is given in~\cite{Georgiou2015arXiv}.

\section{Resource Analysis with CiaoPP} 
\label{sec:ciaopp-analysis}

In order to perform the global energy consumption analysis,
our approach leverages the \ciaopp
tool~\cite{ciaopp-sas03-journal-scp}, the preprocessor of the \ciao{}
programming
environment~\cite{hermenegildo11:ciao-design-tplp-short}. \ciaopp
includes a global static analyzer which is parametric with respect to
resources and type of approximation (lower and upper
bounds)~\cite{resource-iclp07,plai-resources-iclp14}. The framework
can be instantiated to infer bounds on a very general notion of
resources, which we adapt in our case to the inference of energy
consumption.
% Extension
As mentioned before, the resource analysis in \ciaopp works on the
intermediate block-based representation language, which we have called
\hcir in this paper.  Each block is represented as a Horn Clause, so
that, in essence, the \hcir is a pure Horn clause subset (pure logic
programming subset) of the \ciao programming language.
% End
In \ciaopp, a resource is a user-defined \emph{counter} representing a
(numerical) non-functional global property, such as execution time,
execution steps, number of bits sent or received by an application
over a socket, number of calls to a predicate, number of accesses to a
database, etc.
The instantiation of the framework for energy consumption (or any
other resource) is done by means of an assertion language that allows
the user to define resources and other parameters of the analysis by
means of assertions. Such assertions are used to assign basic resource
usage functions to elementary operations and certain program
constructs of the base language, thus expressing how the execution of
such operations and constructs affects the usage of a particular
resource. 
The resource consumption provided can be a constant or a function of
some input data values or sizes.  The same mechanism is used as well
to provide resource consumption information for procedures from
libraries or external code when code is not available or to increase
the precision of the analysis.

For example, in order to instantiate the \ciaopp general analysis
framework for estimating bounds on energy consumption, we start by
defining the identifier (``counter'') associated to the energy
consumption resource, through the following \ciao declaration:
\begin{lstlisting}[language=ciao, basicstyle=\small\ttfamily,]
:- resource energy.
\end{lstlisting}

We then provide assertions for each \hcir block expressing the energy
consumed by the corresponding \llvmir block, determined from the
energy model, as explained in Section~\ref{sec:energy-model}.  Based
on this information, the global static analysis can then infer bounds
on the resource usage of the whole program (as well as procedures and
functions in it) as functions of input data sizes. A full description
of how this is done can be found in~\cite{plai-resources-iclp14}.

Consider the example in Figure~\ref{llvm2hcirblock} (right). Let $P_e$ denote
the energy consumption function for a predicate $P$ in the \hcir
representation (set of blocks with the same name). Let $c_b$ represent
the energy cost of an \llvmir block $b$.
Then, the inferred equations for the \hcir blocks in
Figure~\ref{llvm2hcirblock} (right) are:

$$
\begin{array}{rcl}
alloca_{e}(N, Arr) = c_{alloca} + looptest_{e}(N, Arr) \\
\end{array}
$$
\vspace*{-5mm}
$$
\begin{array}{rcl}
looptest_{e}(N, Arr) = c_{looptest} + loopbody\_loopend_{e}(0 \neq N, N, Arr)\\
\end{array}
$$
\vspace*{-5mm}
$$
\begin{array}{rcl}
loopbody\_loopend_{e}(B, N, Arr)  =

\left\{
\begin{array}{lr}
looptest_{e}(N-1, Arr) \ \ \text{if } B \text{ is {\tt true}} \\
     + \ c_{loopbody} \\ \\

c_{loopend} \ \quad\quad\quad\quad \text{if } B \text{ is {\tt false}}
\end{array}
\right.
\end{array}
$$

If we assume (for simplicity of exposition) that each \llvmir block
has unitary cost, i.e., $c_b = 1$ for all \llvmir blocks $b$, solving
the above recurrence equations, we obtain the energy consumed by
\texttt{alloca} as a function of its input data size ($N$):
\begin{center}
$alloca_{e}(N, Arr) = 2 \times N + 3$
\end{center}

Note that using average energy values in the model implies that the energy
function for the whole program inferred by the upper-bound resource
analysis is an approximation of the actual upper bound (possibly below
it).  Thus, theoretically, to ensure that the analysis infers an
upper bound, we need to use upper bounds as well in the energy models.
This is not a trivial task as the worst case energy consumption depends on the
data processed, is likely to be different for different instructions, and
unlikely to occur frequently in subsequent instructions. 
A first investigation into the effect of different data on the energy consumption of individual instructions, instruction sequences and full programs is presented in~\cite{pallister2015data}. 
A refinement of the energy model to capture upper bounds for individual
instructions, or a selected subset of instructions, is currently being
investigated, extending the first experiments into the impact of data into
worst case energy consumption at instruction level as described in
Section~5.5~of~\cite{Kerrison13}.

\section{Experimental Evaluation}
\label{sec:experiments}

We have performed an experimental evaluation of our techniques on a
number of selected benchmarks. Power measurement data was collected
for the XCore platform by using appropriately instrumented power
supplies, a power-sense chip, and an embedded system for controlling
the measurements and collecting the power data.  
Details about the power monitoring setup used to run our benchmarks
and measure their energy consumption can be found in~\cite{Kerrison13}. 
The main goal of our
experiments was to shed light on the trade-offs implied by performing
the analysis at the ISA \level (without using complex mechanisms for
propagating type information and representing memory) and at the \llvm
level using models defined at the ISA \level together with a mapping
mechanism.

There are two groups of benchmarks that we have used in our
experimental study.
The first group is composed of four small recursive numerical programs
that
have a variety of user defined functions, arguments, and calling
patterns (first four benchmarks in Table~\ref{tab:isa-llvm-comparison}). These benchmarks only operate over primitive data types and
do not involve any structured types.
The second group of benchmarks (the last five benchmarks in
Table~\ref{tab:isa-llvm-comparison}) differs from the first group in
the sense that they all involve structured types. These are recursive
or iterative.

% Extension
The second group of benchmarks includes two filter benchmarks namely
\textit{Biquad} and \textit{Finite Impulse Response (FIR)}. A filter
program attenuates or amplifies one specific frequency range of a
given input signal.  The \texttt{fir(N)} benchmark computes the
inner-product of two vectors: a vector of input samples, and a vector
of coefficients. The more coefficients, the higher the fidelity, and
the lower the frequencies that can be filtered. On the other hand, the
Biquad benchmark is an equaliser running Biquad filtering. An
equaliser takes a signal and attenuates/amplifies different frequency
bands. In the case of an audio signal, such as in a speaker or
microphone, this corrects the frequency response. The
\texttt{biquad(N)} benchmark uses a cascade of Biquad filters where
each filter attenuates or amplifies one specific frequency range.  The
energy consumed depends on the number of banks \texttt{N}, typically
between 3 and 30 for an audio equaliser. A higher number of banks
enables a designer to create more precise frequency response curves.
% End

None of the XC benchmarks contain any assertions that provide
information to help the analyzer.
Table~\ref{tab:isa-llvm-comparison-full} shows detailed experimental results.
Column \textbf{SA energy function} shows the energy consumption
functions, which depend on input data sizes, inferred for each program
by the static analyses performed at the ISA and \llvmir \levels
(denoted with subscripts $isa$ and $llvm$ respectively). We can see
that the analysis is able to infer different kinds of functions
(polynomial, exponential, etc.). Column \textbf{HW} shows the actual
energy consumption in nano-joules measured on the hardware
corresponding to the execution of the programs with input data of
different sizes (shown in column \textbf{Input Data Size}). 
\textbf{Estimated} presents the energy consumption estimated by static
analysis. This is obtained by evaluating the functions in column \textbf{SA
  energy function} for the input data sizes in column \textbf{Input
  Data Size}. The value N/A in such column means that the analysis has
not been able to infer any useful 
energy consumption function and, thus, no
estimated value is obtained. Column \textbf{Err vs.\ HW} shows
the error of the values estimated by the static analysis with respect
to the actual energy consumption measured on the hardware, calculated as follows:
$\textbf{Err vs.\ HW}= (\frac{\textbf{LLVM} (or \ \textbf{ISA}) - \textbf{HW}}{\textbf{HW}} \times 100 ) \%$.
 Finally, the last column shows the ratio between the estimations of
 the analysis at the ISA and \llvmir \levels. 

\begin{table}
\begin{minipage}{\textwidth}  
\centering
\begin{tabular}{|>{\raggedleft}p{35mm}|>{\raggedleft}p{13mm}|>{\raggedleft}p{15mm}|>{\raggedleft}p{15mm}|>{\raggedleft}p{15mm}|>{\raggedleft}p{7mm}|>{\raggedleft\arraybackslash}p{7mm}|p{6mm}|}
\hline
\textbf{SA energy} & {\textbf{Input}} & \textbf{HW
(nJ)} & \multicolumn{2}{c|}{\textbf{Estimated (nJ)}} &

\multicolumn{2}{c|}{\textbf{Err vs. HW\%}} &
\textbf{isa}/\\ \cline{4-7}

\textbf{function (nJ)} & \textbf{Size} &  & \textbf{llvm} &

\textbf{isa} & \textbf{llvm} & \textbf{isa} & \textbf{llvm}  \\
\hline \hline

 \raggedright$Fact_{isa}(N)$=
  & N=8 & 227  & 237  & 212  & 4.6 & -6.4 & 0.9\\ \cline{2-8}

 $ 24.26 \ N + 18.43 $
 & N=16 & 426  & 453  & 406  & 6.5 & -4.5 &0.9\\ \cline{2-8}

 \raggedright$Fact_{llvm}(N)$=
 & N=32  & 824  & 886  & 794  & 7.6 & -3.5 & 0.9\\ \cline{2-8}

  $ 27.03 \ N + 21.28 $
 & N=64 & 1690 & 1751 & 1571 & 3.6 & -7.0 & 0.9\\ \hline

\raggedright$Fib_{isa}(N)$\footnote{\label{fibNote}It uses mathematical functions $fib$ and $lucas$, a function expansion would yield:\\ 
 $Fib_{isa}(N)$=$34.87\times 1.62^N+10.8\times(-0.62)^N-30$\\
 $Fib_{llvm}(N)$=$40.13\times 1.62^N+11.1\times(-0.62)^N-35.65$
 }=$26.88fib(N)$
 & N=2  &  75 &  74 & 65 & -1.16 & -12.5 & 0.89
\\ \cline{2-8}

$+22.85 \ lucas(N)$\footnote{$Lucas(n)$ satisfy the recurrence relation $L_n=L_{n-1}+L_{n-2}$ with $L_1=1, L_2=3$}$-30.04$
 & N=4  &  219 &  241 & 210 & 10 &-4.1& 0.87 \\ \cline{2-8}

& N=8  & 1615   & 1853 & 1608 &14.75&-0.4& 0.87  \\ \cline{2-8}

 \raggedright$Fib_{llvm}(N)^a$=$32.5fib(N)$
& N=15 & $47\times 10^3$ & $54\times 10^3$ & $47\times 10^3$ &16.47&1.2& 0.87\\ \cline{2-8}

 $+25.6 \ lucas(N)^b-35.65$
 & N=26 & $9.30\times10^6$ & $10.9\times 10^6$ & $9.5\times 10^6$ &17.3&1.74& 0.87 \\ \hline 

 \raggedright$Sqr_{isa}(N)$=
 & N=9  & 1242 &  1302 & 1148  & 4.8 &-7.5  &
0.88\\ \cline{2-8}

 $8.6 \ N^2 + 48.7 \ N + 15.6$ 
  & N=27 & 8135 &  8734 & 7579 & 7.4 & -6.8 & 0.87 \\

\cline{2-8}

 & N=73 & $52\times 10^3$ & $57\times 10^3$ & $49\times 10^3$ & 8.5 & -6.5  & 0.86 \\ \cline{2-8}

 \raggedright$Sqr_{llvm}(N)$= 
 & N=144 & $19.7\times 10^4$ &  $21.4\times 10^4$ & $18.4\times 10^4$ & 8.89 & -6.4 & 0.86 \\ \cline{2-8}

   $10 \ N^2 + 53 \ N + 15.6$ 
& N=234  & $51\times 10^4$ & $56\times 10^4$ & $48\times 10^4$ & 9.61 & -5.86 & 0.86 \\ \cline{2-8}
  
& N=360  & $11.89\times 10^5$ & $13\times 10^5$ & $11.2\times 10^5$ & 10.49 & -5.16 & 0.86 \\ \hline

 & N=3 & 326 & 344 & 3.6 & 5.7 & -6.0 & 0.89 \\ \cline{2-8}

\raggedright$PowerOfTwo_{isa}(N)$=
 & N=6  & 2729 & 2965 & 2631 & 8.7 &3.6 & 0.89 \\ \cline{2-8}

 $41.5\times 2^{N}-25.9 $
  & N=9 & $21.9\times 10^3$ & $23.9\times 10^3$ & $21.2\times 10^3$ &9 & 3.3 & 0.89 \\ \cline{2-8}

 \raggedright$PowerOfTwo_{llvm}(N)$ = 
 & N=12 & $17.57\times 10^4$ & $19.1\times 10^4$ & $17\times 10^4$ & 9 & -3.3 & 0.89 \\ \cline{2-8}
$ 46.8 \times 2^{N} -29.9 $ 
 & N=15 & $13.8\times 10^5$ & $15.3\times 10^5$ & $13.6\times 10^5$ & 11 & -1.5 & 0.89 \\ \hline
 \hline
 
 & N=57 & 1138 & 1179 & N/A & 3.60 & N/A & N/A \\ \cline{2-8}

\raggedright$reverse_{llvm}(N)$=
 & N=160  & 3125 & 3185 & N/A & 1.91 & N/A & N/A\\ \cline{2-8}

 $19.47 \ N + 69.33 $
  & N=320 & 6189 & 6301 & N/A &1.82 & N/A & N/A\\ \cline{2-8}

 & N=720 & 13848 & 14092 & N/A & 1.76 & N/A & N/A\\ \cline{2-8}

 & N=1280 & 24634 & 24998 & N/A & 1.48 & N/A & N/A\\ \hline

 & N=5 & 7453 & 7569 & N/A & -2 & N/A & N/A \\ \cline{2-8}

\raggedright$matmult_{llvm}(N)$=
 & N=15  & $15.79\times 10^4$ & $15.9\times 10^4$ & N/A & 1.03 & N/A & N/A\\ \cline{2-8}

 $42.47 \ N^3 + 68.85 \ N^2+$
  & N=20 & $36.29\times 10^4$ & $36.8\times 10^4$ & N/A &1.51 & N/A & N/A\\ \cline{2-8}

 $49.9 \ N + 24.22 $& N=25 & $69.56\times 10^4$ & $70.8\times 10^4$ & N/A & 1.77 & N/A & N/A\\ \cline{2-8}

 & N=31 & $13.07\times 10^5$ & $13.3\times 10^5$ & N/A & 1.98 & N/A & N/A\\ \hline

 & N=131; M=69  & $14.5\times 10^3$ & $13.2\times 10^3$ & N/A & 8.65 & N/A & N/A\\ \cline{2-8}

 \raggedright$concat_{llvm}(N,M)$= 
  & N=170; M=182 & $25.44\times 10^3$ & $23.3\times 10^3$ & N/A &8.60 & N/A & N/A\\ \cline{2-8}

 $65.7 \ N + 65.7 \ M + 137 $
 & N=188; M=2 & $13.8\times 10^3$ & $12.6\times 10^3$ & N/A & 8.59 & N/A & N/A\\ \cline{2-8}

 & N=13; M=134 & $10.7\times 10^3$ & $9.79\times 10^3$ & N/A & 8.74 & N/A & N/A\\ \hline

\raggedright$biquad_{llvm}(N)$= 	  & N=5  & 871 & 836 & N/A & -4 & N/A & N/A\\  \cline{2-8}
$157 \ N + 51.7 $	  & N=7  & 1187 & 1151 & N/A & -3.1 & N/A & N/A\\ \cline{2-8}
					  & N=10  & 1660 & 1622 & N/A & -2.31 & N/A & N/A\\ \cline{2-8}
					  & N=14  & 2290 & 2250 & N/A & -1.75 & N/A & N/A\\ \hline

\raggedright$fir_{llvm}(N)$=   & N=85  & 2999 & 2839 & N/A & -5.3 & N/A & N/A\\  \cline{2-8}
$31.8 \ N + 137 $ & N=97  & 3404 & 3221 & N/A & -5.37 & N/A & N/A\\  \cline{2-8}
				   & N=109  & 3812 & 3602 & N/A & -5.5 & N/A & N/A\\  \cline{2-8}
				   & N=121  & 4227 & 3984 & N/A & -5.7 & N/A & N/A\\  \hline

\end{tabular}
\caption{Comparison of the accuracy of energy analyses at the \llvmir and ISA \levels.}
\label{tab:isa-llvm-comparison-full}
\end{minipage}
\end{table}

Table~\ref{tab:isa-llvm-comparison} shows a summary of results. The
first two columns show the name and short description of the
benchmarks. The columns under \textbf{Err vs. HW} show the average
error obtained from the values given in
Table~\ref{tab:isa-llvm-comparison-full} for different input data
sizes.
The last row of the table shows the average error over the number of
benchmarks analyzed at each \level.

\begin{table}[ht]
\centering

\begin{tabular}{|l|l|r|r|c|}
\hline 
\textbf{Program} &\textbf{Description} & \multicolumn{2}{c|}{\textbf{Err vs. HW}} & \textbf{isa/}
\\  \cline{3-4}
&&\textbf{llvm}&\textbf{isa}&\textbf{llvm}
\\ \hline \hline 

\texttt{fact(N)}  &  Calculates N!      & 5.6\% & 5.3\% & 0.89
\\  \hline 
\texttt{fibonacci(N)}  &   Nth Fibonacci number   & 11.9\%& 4\%& 0.87
\\ \hline 
\texttt{sqr(N)}     &     Computes $N^2$ performing additions    & 9.3\%&3.1\% & 0.86
\\ \hline 
\texttt{pow\_of\_two(N)}   &   Calculates $2^N$ without multiplication & 9.4\% &3.3\% & 0.89
\\ \hline 
\texttt{\textbf{Average}} & & \textbf{9\%} & \textbf{3.9\%} & \textbf{0.92}
\\ \hline \hline
\texttt{reverse(N, M)} &  Reverses an array  & 2.18\%& N/A& N/A
\\ \hline 
\texttt{concat(N, M)}  &  Concatenation of arrays   &8.71\%&N/A& N/A
\\ \hline 
\texttt{matmult(N, M)}  &  Matrix multiplication  & 1.47\% & N/A & N/A
\\ \hline 
\texttt{fir(N)} & Finite Impulse Response filter & 5.47\% & N/A & N/A
\\ 
\hline
\texttt{biquad(N)} & Biquad equaliser & 3.70\% & N/A & N/A
\\ 
\hline 
\texttt{\textbf{Average}} & & \textbf{3.0\%} & \textbf{N/A} & \textbf{N/A}
\\ \hline
\hline 
\texttt{\textbf{Overall average}} & & \textbf{6.4\%} & \textbf{3.9\%} & \textbf{0.92}
\\ \hline
\end{tabular}
\caption{\llvmir- vs.\ ISA-\level analysis accuracy.}
\label{tab:isa-llvm-comparison}
\tabend
\tabend
\end{table}

The experimental results show that:
\begin{itemize}

\item For the benchmarks in the first group, both
  the ISA- and \llvmir-\level analyses are able to infer useful energy
  consumption functions.  On average, the analysis performed at either
  level is reasonably accurate and the relative error between the two
  analyses at different \levels is small.  ISA-\level estimations are
  slightly more accurate than the ones at the \llvmir \level (3.9\%
  vs.\ 9\% 
  error on average with
  respect to the actual energy consumption measured on the hardware, 
  respectively). This is because the ISA-\level analysis uses very
  accurate energy models, obtained from measuring directly at the ISA
  \level, whereas at the \llvmir \level, such ISA-\level model needs
  to be propagated up to the \llvmir \level using (approximated)
  mapping information. This causes a slight loss of accuracy.

\item For the second group of benchmarks, the ISA \level analysis
is not able to infer useful 
energy functions.  This is due to the fact that
significant program structure and data type/shape information is lost
due to lower-level representations, which sometimes makes the analysis
at the ISA \level very difficult or impossible.
In order to overcome this limitation and improve analysis accuracy,
significantly more complex techniques for recovering type information
and representing
memory in the \hcir would be needed.  In contrast, type/shape
information is preserved at the \llvmir \level, which allows analyzing
programs using data structures (e.g., arrays). In particular, all the
benchmarks in the second group are analyzed at the \llvmir \level with
reasonable accuracy (3\%
error on average). 
In this sense, the \llvmir-\level analysis is more powerful than the
one at the ISA \level.  
The analysis is also reasonably efficient, with analysis times of
about 5 to 6 seconds on average, despite the naive implementation of
the interface with external recurrence equation solvers, which can be
improved significantly. The scalability of the analysis follows from
the fact that it is compositional and can be performed in a modular
way, making use of the \ciao assertion language to store results of
previously analyzed modules.

\end{itemize}

\section{Related Work}
\label{sec:related-work}

Few papers can be found in the literature focusing on static analysis
of energy consumption. 
% Extension
As mentioned before, the approach presented in this paper builds on
our previously developed analysis of XC
programs~\cite{isa-energy-lopstr13-final} based on transforming the
corresponding ISA code into a Horn Clause representation that is
supplied, together with an ISA-\level energy model, to the
\ciaopp~\cite{ciaopp-sas03-journal-scp} resource analyzer.  In this
work we have increased the power of the analysis by transforming and
analyzing the corresponding \llvmir, and using techniques for
reflecting the ISA-\level energy model upwards to the \llvmir \level.
We also offer novel results supported by our experimental study that
shed light on the trade-offs implied by performing the analysis at
each of these two \levels. Our approach now enables the analysis of a
wider range of benchmarks.  We obtained promising results for a good
number of benchmarks for which~\cite{isa-energy-lopstr13-final} was
not able to produce useful energy functions.
% End
A similar approach was proposed for upper-bound energy analysis of
Java bytecode programs in~\cite{NMHLFM08}, where the Jimple (a typed
three-address code) representation of Java bytecode was transformed
into Horn Clauses, and a simple energy model at the Java bytecode
\level~\cite{LL07} was used.  However, this work did not compare the
results with actual, measured energy consumption.

In all the approaches mentioned above, instantiations for energy
consumption of general resource analyzers are used,
namely~\cite{resource-iclp07} in~\cite{NMHLFM08}
and~\cite{isa-energy-lopstr13-final},
and~\cite{plai-resources-iclp14} in this paper. Such resource
analyzers are based on setting up and solving recurrence equations, an
approach proposed by Wegbreit~\cite{DBLP:journals/cacm/Wegbreit75}
that has been developed significantly in subsequent
work~\cite{Rosendahl89,granularity,low-bounds-ilps97,vh-03,resource-iclp07,AlbertAGP11a,plai-resources-iclp14}.
Other approaches to static analysis
based on the transformation of the analyzed code into another
(intermediate) representation have been proposed for analyzing
low-level languages~\cite{HenriksenG06} and Java (by means of a
transformation into Java
bytecode)~\cite{jvm-cost-esop}. In~\cite{jvm-cost-esop}, cost
relations are inferred directly for these bytecode programs, whereas
in~\cite{NMHLFM08} the bytecode is first transformed into Horn
Clauses.
The general resource analyzer in~\cite{resource-iclp07} was also
instantiated in~\cite{estim-exec-time-ppdp08} for the estimation of
execution times of logic programs running on a bytecode-based abstract
machine. The approach used timing models at the bytecode instruction
level, for each particular platform,
and program-specific mappings to lift such models up to the Horn
Clause level, at which the analysis was performed. 
% Extension
The timing model was automatically produced in a one-time,
program-independent profiling stage by using a set of synthetic
calibration programs and setting up a system of linear equations.
% end

By contrast to the generic approach based on \ciaopp, an approach operating directly
on the \llvmir representation is explored in~\cite{grech15}. Though relying on similar analysis
techniques, the approach can be integrated more directly in the LLVM
toolchain and is in principle applicable to any languages targeting
this toolchain. The approach uses the same \llvmir energy model and mapping technique
as the one applied in this paper.

% Extension
There exist other approaches to cost analysis such as those using
dependent types \cite{DBLP:journals/toplas/0002AH12}, SMT solvers
\cite{Alonso2012}, or \emph{size change abstraction}
\cite{DBLP:journals/corr/abs-1203-5303}
% End

A number of static analyses are also aimed at worst case execution time
(WCET), usually for imperative languages in different application
domains (see e.g.,~\cite{DBLP:journals/tecs/WilhelmEEHTWBFHMMPPSS08} and its
references). 
The worst-case analysis presented
in~\cite{DBLP:conf/rtas/JayaseelanML06}, which is not based on
recurrence equation solving, distinguishes instruction-specific (not
proportional to time, but to data) from pipeline-specific (roughly
proportional to time) energy consumption.
However, in contrast to the work presented here and
in~\cite{estim-exec-time-ppdp08}, these worst case analysis 
methods do not infer cost functions on input data sizes but rather
absolute maximum values, and they generally require the
manual annotation of loops to express an upper-bound on the number of
iterations.
An alternative approach to WCET was presented
in~\cite{Herrmann-WCET-2007}.  It is based on the idea of
amortisation, which allows to infer more accurate yet safe upper
bounds by averaging the worst execution time of operations over time.
It was applied to a functional language, but the approach is in
principle generally applicable.
A timing analysis based on game-theoretic learning was presented
in~\cite{DBLP:conf/tacas/SeshiaK11}. The approach combines static
analysis to find a set of basic paths which are then tested. In
principle, such approach could be adapted to infer energy usage. Its
main advantage is that this analysis can infer distributions on time,
not only average values.

\section{Conclusions and Future Work}
\label{sec:conclusions}

We have presented techniques for 
extending  
to the \llvmir \level our
tool chain for estimating energy consumption as functions on program
input data sizes. The approach uses a mapping technique that leverages
the existing debugging mechanisms in the XMOS XCore compiler tool
chain to propagate an ISA-\level energy model to the \llvmir
\level. A new transformation constructs a block representation that 
is supplied, together with the propagated energy values, to a parametric
resource analyzer that infers the program energy cost as functions on the
input data sizes.

Our results suggest that performing the static analysis at the \llvmir
\level is a reasonable compromise, since 1) \llvmir is close enough to
the source code \level to preserve most of the program information
needed by the static analysis, and 2) the \llvmir is close enough to
the ISA \level to allow the propagation of the ISA energy model up to
the \llvmir \level without significant loss of accuracy for the
examples studied.  Our experiments are based on single-threaded
programs.
We also have focused on the study of the energy consumption due to
computation, so that we have not tested programs where storage and
networking is important. However, this could potentially be done in
future work, by using the \ciaopp static analysis, which already
infers bounds on data sizes, and combining such information with
appropriate energy models of communication and storage.
% Extension
Although the analysis infers sound bound representations in the form
of recurrence equations, sometimes the external solvers it uses are
not able to find closed form functions for such equations. This is a
limitation in applications where such closed forms are needed.
Techniques to address such limitation are included in our plans for
future work. Our static analysis will also benefit from any
improvement of the Computer Algebra Systems used for solving
recurrence equations.
% End
  
It remains to be seen whether the results would carry over to other
classes of programs, such as multi-threaded programs and programs
where timing is more important. In this sense our results are
preliminary, yet they are promising enough to continue research into
analysis at \llvmir \level and into ISA-\llvmir energy mapping
techniques to enable the analysis of a wider class of programs,
especially multi-threaded programs.

\section*{Acknowledgements}

This research has received funding from the European Union 7th
Framework Program agreement no 318337, ENTRA, Spanish MINECO
TIN'12-39391 \emph{StrongSoft} project, and the Madrid M141047003
\emph{N-GREENS} program.

\bibliographystyle{abbrv}

\end{document}